\documentclass[twocolumn,superscriptaddress, english,prl]{revtex4-1}
\usepackage[colorlinks=true,urlcolor=blue,citecolor=blue,linkcolor=blue]{hyperref} 
\usepackage[T1]{fontenc}
\usepackage[latin9]{inputenc}
\usepackage{amssymb}
\usepackage{graphicx}
\usepackage{amsmath,color}
\usepackage{mathrsfs}
\usepackage{float}
\usepackage{indentfirst}
\usepackage{braket}

\makeatletter

\def\journal #1, #2, #3, 1#4#5#6{{\sl #1~}{\bf #2}, #3 (1#4#5#6) }

\def\eqa{\begin{eqnarray}}
\def\eea{\end{eqnarray}}
\newcommand{\eq}{\begin{equation}}
\newcommand{\ee}{\end{equation}}

\newcommand{\Eq}[1]{Eq.~(\ref{#1})}

\makeatother

\usepackage{babel}

\begin{document}

\title{Accelerate Monte Carlo Simulations with Restricted Boltzmann Machines}

\author{Li Huang}
\affiliation{Science and Technology on Surface Physics and Chemistry Laboratory, P.O. Box 9-35, Jiangyou 621908, China}
\author{Lei Wang}
\email{wanglei@iphy.ac.cn}
\affiliation{Beijing National Lab for Condensed Matter Physics and Institute
of Physics, Chinese Academy of Sciences, Beijing 100190, China }

\begin{abstract}
Despite their exceptional flexibility and popularity, the Monte Carlo methods often suffer from slow mixing times for challenging statistical physics problems. We present a general strategy to overcome this difficulty by adopting ideas and techniques from the machine learning community. We fit the unnormalized probability of the physical model to a feedforward neural network and reinterpret the architecture as a restricted Boltzmann machine. Then, exploiting its feature detection ability, we utilize the restricted Boltzmann machine for efficient Monte Carlo updates and to speed up the simulation of the original physical system. We implement these ideas for the Falicov-Kimball model and demonstrate improved acceptance ratio and autocorrelation time near the phase transition point. 
\end{abstract}
\maketitle

Monte Carlo method is one of the most flexible and powerful methods for studying many-body systems~\cite{Newman:1999fd, Krauth:2006vv}. Its application ranges from the physical sciences~\cite{gubernatis2003monte} including condensed matter physics~\cite{2001RvMP...73...33F, gubernatis2016quantum}, nuclear matter~\cite{Carlson:2015fu} and particle physics~\cite{Fodor:2012cea} all the way to the biological and social sciences~\cite{manly2006randomization, mode2011applications, glasserman2003monte}. Monte Carlo methods randomly sample configurations and obtain the answer as a statistical average. However, because of the configuration spaces are exceptionally large for many-body systems, it is typically impossible to perform direct sampling, therefore one resorts to the Markov chain random walk approach to explore the configuration space. In this case, one only needs to know the relative ratio between the probabilities of two configurations.  

Designing efficient strategies to explore the configuration space efficiently is at the heart of Markov chain Monte Carlo algorithms. This is, however, a challenging endeavor. Not even mentioning the fundamentally difficult case of glassy energy landscapes, naive Monte Carlo samplings are usually painfully slow close to the phase transitions. These drawbacks motivated noteworthy algorithmic developments in the past decades~\cite{duane1987hybrid, 1987PhRvL..58...86S, Wolff:1989iy, Prokofev:1998tc, 2001PhRvE..64e6101W, PhysRevLett.70.875}. In essence, those algorithms exploit various physical aspects of the problem for efficient Monte Carlo updates. It is however difficult to devise a general strategy to guide optimal Monte Carlo algorithm design. 

We address these difficulties in a general setting with insights from the machine learning. Recently, there has been a rising interest in applying machine learning approaches to many-body physics problems. This includes classifying phases of matter~\cite{Carrasquilla:2016wu, Wang:2016tfa, Anonymous:PxlD92x3, Anonymous:jxjVcLI0,  2016arXiv160909087T}, using the neural networks as variational wave functions~\cite{Carleo:2016vp, 2016arXiv160909060D}, fitting the density functionals~\cite{Snyder:2012da, Li:2016vh,Brockherde:2016uz}, and solving inverse problems in quantum many-body physics~\cite{Arsenault:2014em, Arsenault:2015wd}. 

In this paper, we propose a general way to accelerate Monte Carlo simulations of statistical physics problems. We present two algorithmic innovations: a simple \emph{supervised learning} approach to train the restricted Boltzmann machine (RBM)~\cite{Smolensky:1986va,Hinton:2002ic} as a proxy of the  physical distributions, and an efficient Monte Carlo sampling strategy that exploits the latent structure of the RBM. The RBM is a building block for deep learning and plays an important role in its the recent renaissance~\cite{Hinton:2006kc}. The significance of using the RBM in the Monte Carlo simulations is that it automatically identifies relevant features (such as correlations and collective modes) in the physics model and proposes updates correspondingly with high acceptance rates and low autocorrelations. This approach makes better use of the sampled Monte Carlo data because in addition to estimate the physical observables, the RBM builds an adaptive model for the physical probability distribution and guides better explorations. 

\begin{figure}[!t]
  \centering
 \includegraphics[width=\columnwidth]{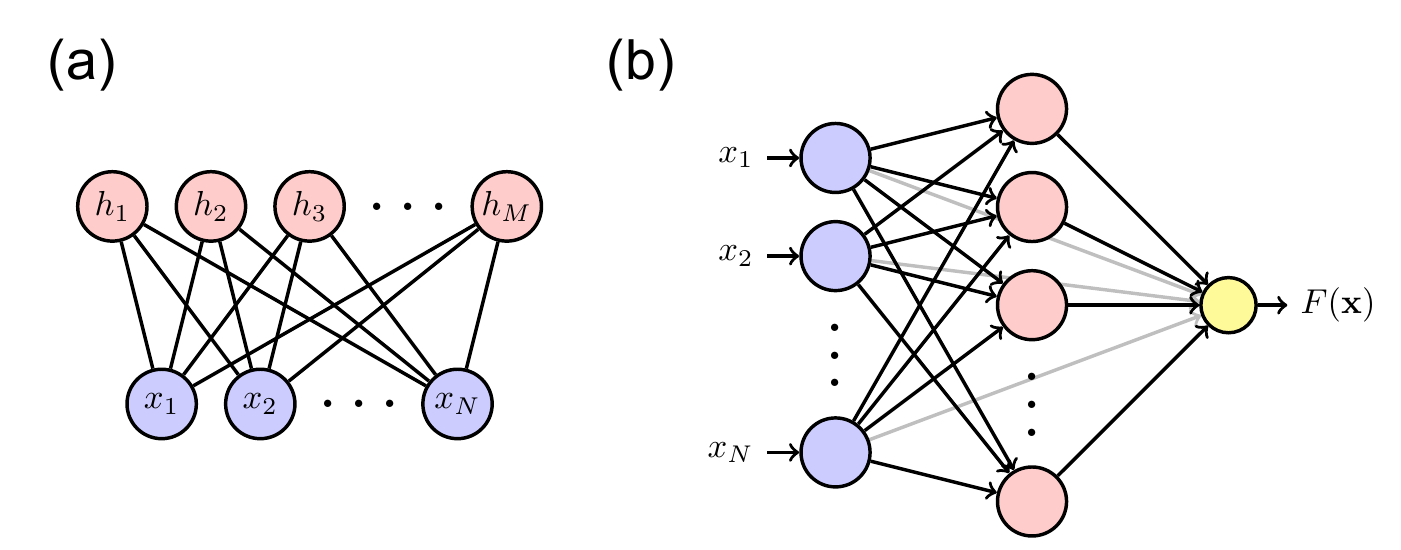}
  \caption{(a) The restricted Boltzmann machine is an energy-based model for the binary stochastic visible and hidden variables. Their probability distribution follow the Boltzmann distribution with the energy function in~\Eq{eq:RBM}. (b) Viewing the RBM as a feedforward neural network which maps the visible variables to the free energy \Eq{eq:FRBM}. The gray arrows represent the first term of \Eq{eq:FRBM}. The red circles are hidden neurons with softplus activation function, corresponding to the second term of \Eq{eq:FRBM}. Fitting $F(\mathbf{x})$ to the log-probability of the physical models (e.g. \Eq{eq:FFK}) determines the weights and biases of the RBM.}
  \label{fig:RBMNN}
\end{figure} 

We illustrate these general ideas using the Falicov-Kimball model~\cite{Falicov:1969ua} as an example. The model describes mobile fermions and localized fermions interact with onsite interactions. The Hamiltonian reads   
\begin{equation}
\hat{H}_\mathrm{FK} = \sum_{i,j} \hat{c}^{\dagger}_{i} \mathcal{K}_{ij} \hat{c}_{j}+  U \sum_{i=1}^{N} \left(\hat{n}_{i}-\frac{1}{2}\right) \left(x_{i}-\frac{1}{2}\right),
\label{eq:FKmodel}
\end{equation}
where $x_{i}\in\{0,1\}$ is a classical binary variable representing the occupation number of the localized fermion at site $i$. $\hat{c}_{i}$ is the fermion annihilation operator and $\hat{n}_{i}\equiv\hat{c}_{i}^{\dagger}\hat{c}_{i}$ is the occupation number operator of the mobile fermion. $\mathcal{K}$ is the kinetic energy matrix of the mobile fermions. In the following, we consider the model on a periodic square lattice with $N$ sites. Thus, $\mathcal{K}_{ij}=-t$ for nearest neighbors and is zero otherwise. The $-1/2$ offsets in \Eq{eq:FKmodel} ensures that both the mobile and localized fermions are half-filled on average. Previous studies show that at $U/t=4$ and temperature $T/t\approx0.15$ the system undergoes a phase transition to the checkerboard density wave (CDW) state~\cite{Maska:2006id,Antipov:2014ij,PhysRevLett.117.146601}. 

Tracing out the mobile fermions, the occupation number of the localized fermions $\mathbf{x}\in\{0,1\}^{N}$ follow the probability distribution $p_\mathrm{FK}(\mathbf{x})= e^{-F_\mathrm{FK}(\mathbf{x})}/Z_\mathrm{FK}$, where $Z_\mathrm{FK}$ is a normalization factor. The negative ``free energy'' reads (omitting an unimportant constant $\beta U N/4$) 
\begin{equation}
-F_\mathrm{FK}(\mathbf{x}) = \frac{\beta U}{2} \sum_{i=1}^{N}x_{i} +\ln\det\left(1+e^{-\beta \mathcal{H}}\right), 
\label{eq:FFK}
\end{equation}
where $\beta=1/T$ is the inverse temperature and $\mathcal{H}_{ij} = \mathcal{K}_{ij} + \delta_{ij}U\left(x_{i}-1/2\right)$. One can diagonalize $\mathcal{H}$ to obtain its eigenvalues $\varepsilon_{i}$ and compute $\sum_{i=1}^{N} \ln (1+e^{-\beta \varepsilon_{i}})$ for the second term of \Eq{eq:FFK}. The case of classical fields coupled to quadratic fermions represents a broad class of physics problems, including the double-exchange model~\cite{Alvarez:2003ex}, the mean-field model for phase fluctuated superconductors~\cite{Mayr:2005jm, Dubi:2007kz} and the Kitaev model after a transformation~\cite{Nasu:2014gk}. Moreover, if one allows imaginary-time dependence in~\Eq{eq:FFK}, it covers an even broader class of condensed matter physics problems such as the Hubbard models~\cite{Blankenbecler:1981vj} and spin-fermion models~\cite{Berg:2012ie}. 

To compute the physical properties of the Falicov-Kimball model (\ref{eq:FKmodel}), one can perform the Monte Carlo sampling of the classical variables $\mathbf{x}$. To this end, one designs an ergodic strategy to update the variables $\mathbf{x}$ and decide whether accept or reject each move. It is sufficient for the  Markov chain to converge to the true distribution if the updates satisfy the detailed balance condition
~\cite{Metropolis:1953in, Hastings:1970aa},  
\begin{equation}
\frac{T(\mathbf{x}\rightarrow{\mathbf{x}'})}{T(\mathbf{x}'\rightarrow \mathbf{x})} \frac{A(\mathbf{x}\rightarrow{\mathbf{x}'})}{A(\mathbf{x}'\rightarrow \mathbf{x})} = \frac{p_\mathrm{FK}(\mathbf{x}')}{p_\mathrm{FK}(\mathbf{x})}, 
\label{eq:dbl}
\end{equation} 
where $T(\mathbf{x}\rightarrow{\mathbf{x}'})$ is the proposal probability of an update and $A(\mathbf{x}\rightarrow{\mathbf{x}'})$ is the acceptance probability of the update. A naive approach would randomly change the classical variables and recompute \Eq{eq:FFK}. To keep the acceptance rate high, one typically applies local updates such as randomly selects a site $i$ and tries to flip the bit $x_{i}\rightarrow1-x_{i}$. In this case the ratio $\frac{T(\mathbf{x}\rightarrow{\mathbf{x}'})}{T(\mathbf{x}'\rightarrow \mathbf{x})}=1$ and the acceptance ratio only depends on the free energy difference of the physical model~\Eq{eq:FFK}. However, this naive approach not only has an unfavorable $\mathcal{O}(N^{4})$ scaling with the system size but also has long autocorrelation times. Several improved update schemes~\cite{Motome:1999ig, 2001NuPhB.596..587A, Alvarez:2005cv,Kumar:2006iea} have been developed by exploiting the specific features of the Falicov-Kimball model~(\ref{eq:FFK}). We next present a general approach to propose efficient Monte Carlo updates ${T(\mathbf{x}\rightarrow{\mathbf{x}'})}$ by training and simulating an RBM. 

The RBM is a classical statistical mechanics system defined by the following energy function
\begin{equation}
E\left(\mathbf{x}, \mathbf{h}\right) = -\sum_{i=1}^{N} a_{i} x_{i} -\sum_{j=1}^{M} b_{j} h_{j} -\sum_{i=1}^{N}\sum_{j=1}^{M} x_{i} W_{ij} h_{j}, 
\label{eq:RBM}
\end{equation}
where $\mathbf{x}\in\{0,1\}^{N}$ and $\mathbf{h}\in\{0,1\}^{M}$ are binary numbers referred as visible and hidden variables. Besides the local biases $a_{i}$ and $b_{j}$, the visible and hidden units are coupled by the weights $W_{ij}$. Crucially, there is no coupling within the visible or hidden units themselves, see Fig.~\ref{fig:RBMNN}(a). The joint probability distribution of the visible and hidden variables follows the Boltzmann distribution $p(\mathbf{x}, \mathbf{h}) = e^{-E(\mathbf{x}, \mathbf{h})}/Z$, where the partition function $Z$ is a normalization factor.  

\begin{figure}[!t]
  \centering
 \includegraphics[width=\columnwidth]{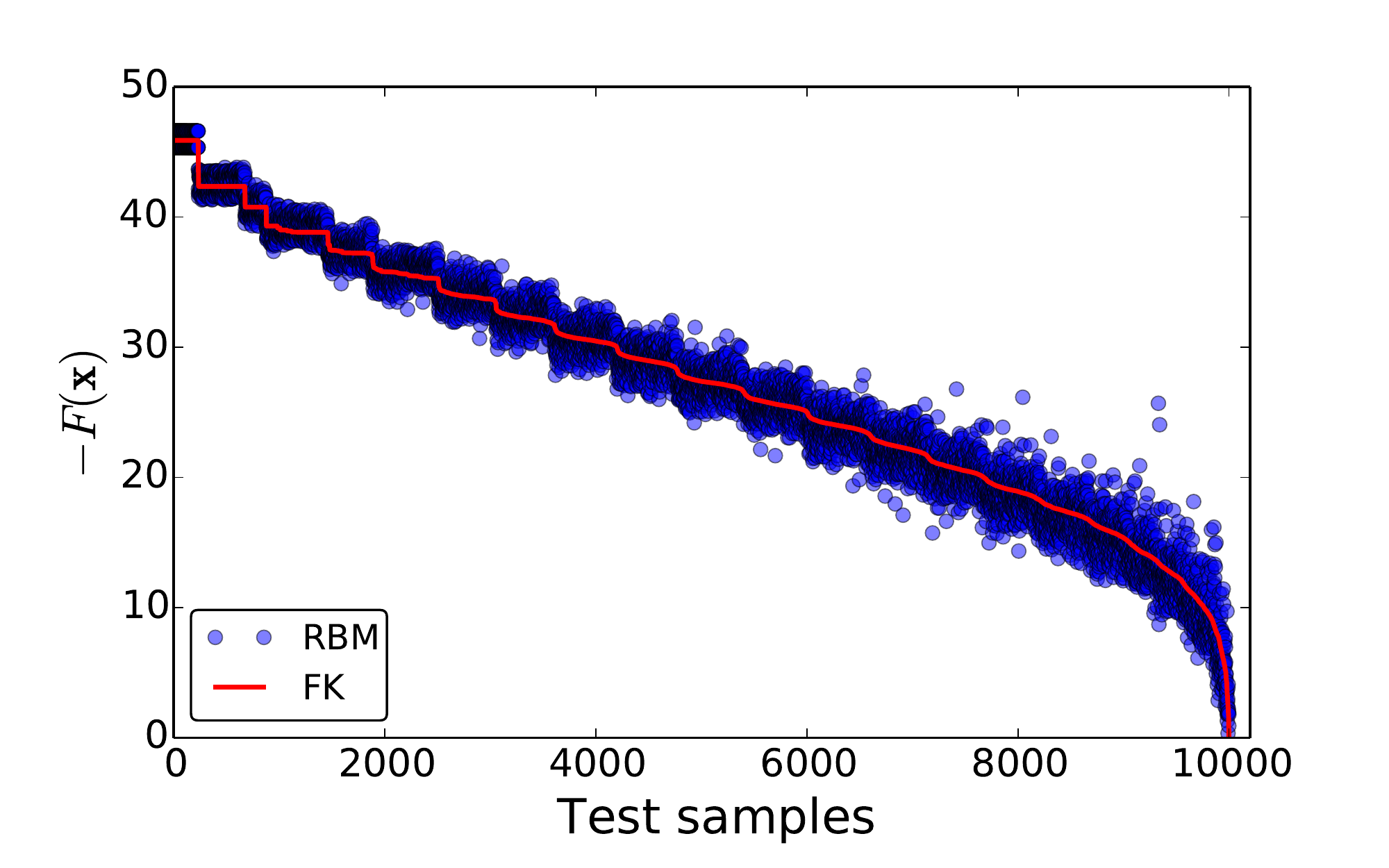}
  \caption{Fitting of the log-probability of the RBM \Eq{eq:FRBM} to the one of the Falicov-Kamball model \Eq{eq:FFK} on a $N=8^{2}$ square lattice.  The parameters are $U/t=4, T/t=0.15$ and we use $M=100$ hidden variables. We offset the training data so that the minimum value is at zero. }
  \label{fig:fit}
\end{figure}

Given sufficiently large number of hidden units, one can tune the parameters of an RBM to let the marginal distribution $p(\mathbf{x}) =\sum_{\mathbf{h}}p(\mathbf{x}, \mathbf{h})$ approximate any discrete distribution~\cite{Freund:1994tu,LeRoux:2008ex}. This training task is similar to solving the ``inverse Ising problem'' in statistical physics~\cite{Albert:2014ih}. The machine learning community has developed practical approaches to train the RBM by minimizing the negative log-likehood $-\ln p(\mathbf{x})$~estimated on the samples drawn from the target distribution~\cite{Hinton:2006dk, Tieleman:2008tj, Hinton:2012ig}. References~\cite{Anonymous:IsmowpLi, Torlai:2016th} uses this approach to train the RBM for the thermal distribution of the classical Ising model. 

\begin{figure}[!t]
  \centering
  \includegraphics[width=\columnwidth]{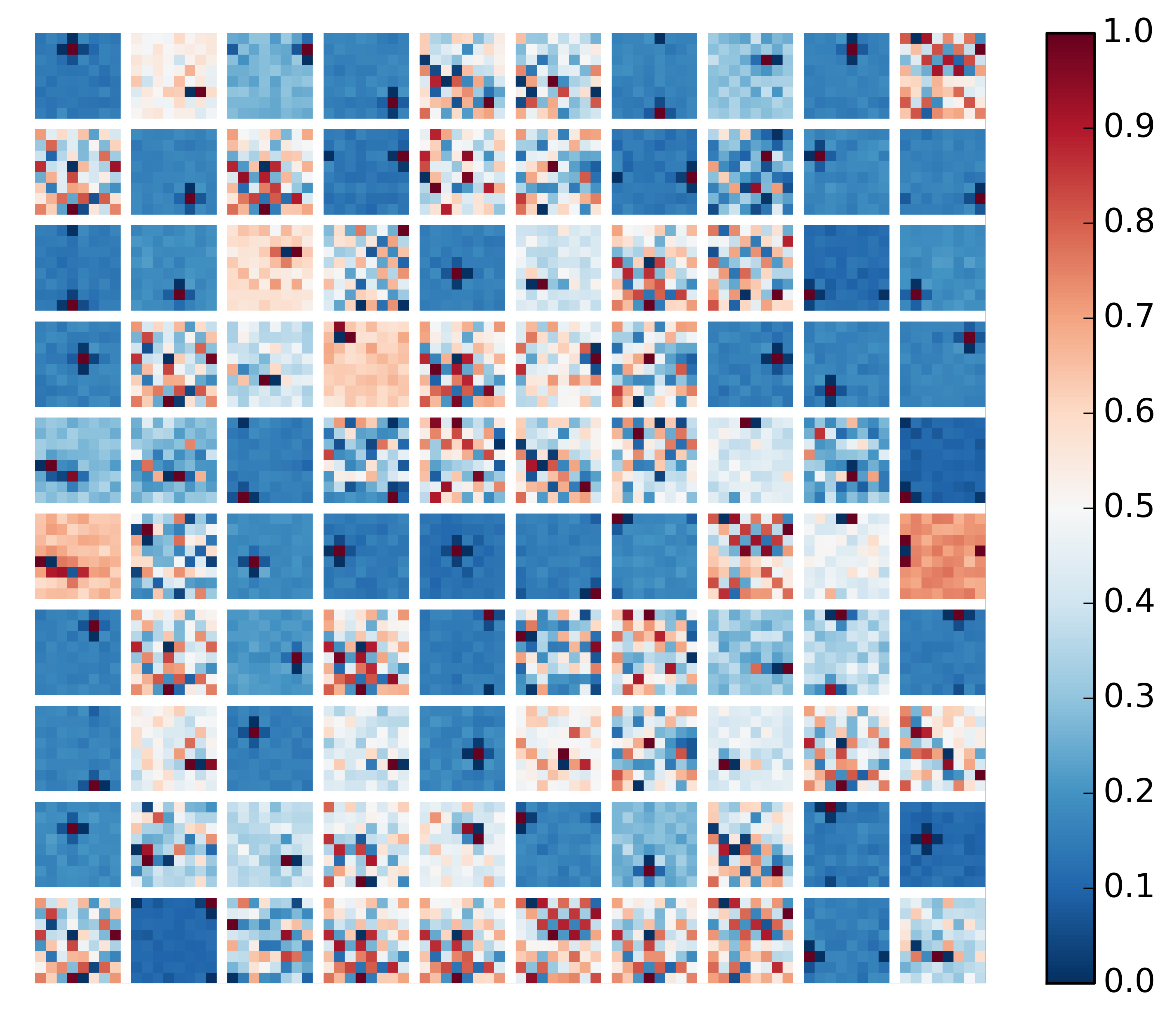}
  \caption{Connection weights $W_{ij}$ of the restricted Boltzmann machine trained for the Falicov-Kimball model (\ref{eq:FKmodel}) on a square lattice with $N=8^{2}$ sites at $U/t=4, T/t=0.15$. Each tile represents the connection weights of a hidden neuron. We scale the weights in the range of $0\sim1$ for visibility.}
  \label{fig:weights}
\end{figure}

However, training the RBM can be even simpler for the statistical physics problems. Two observations are crucial here. First, in addition to the dataset sampled from the target distribution, we do have access to the \emph{unnormalized} probability (i.e. log-probability up to a constant) of the statistical physical problems, e.g. \Eq{eq:FFK}. On the other hand, we can write the marginal distribution of the RBM as $p(\mathbf{x})=e^{-F(\mathbf{x})}/Z$, where the ``free energy'' of the visible variables reads 
\begin{equation}
-F\left(\mathbf{x}\right) = \sum_{i=1}^{N}a_{i} x_{i} +\sum_{j=1}^{M}\ln\left(1+e^{b_{j}+\sum_{i=1}^{N}x_{i}W_{ij}}\right). 
\label{eq:FRBM}
\end{equation}
Viewing the RBM as a functional mapping from $\mathbf{x}$ to $F(\mathbf{x})$ in \Eq{eq:FRBM}~\cite{Anonymous:7L6_jSRi}, it is tempting to employ a \emph{supervised learning} approach to train its parameters 
by fitting the free energy \Eq{eq:FRBM} to the one of the physical model~\Eq{eq:FFK}. Here the second observation is important: since only the relative probability ratio matters in the Markov chain Monte Carlo sampling, one thus only cares about $F(\mathbf{x})$ up to an additive constant. Thus, the intractable partition functions of the RBM and the physical model do not appear in the fitting. The overall constant offset of \Eq{eq:FFK} and \Eq{eq:FRBM} can be chosen at our convenience in the supervised learning. 

We set up a feedforward neural net for~\Eq{eq:FRBM} shown in Fig.~\ref{fig:RBMNN}(b). Its trainable biases and connection weights correspond to the parameters $a_{i}, b_{j}$ and $W_{ij}$  of the RBM. The red hidden neurons activate via the softplus function $f(z)=\ln(1+e^{z})$. The yellow output neuron sums up the outputs of the hidden neurons and the results coming directly from the input neurons for the final result. Physically, the mobile fermions induce effective interactions between the localized fermions. By training an RBM, we represent the fermionic environment by the classical binary fields represented by the hidden neurons. This supervised training approach is significantly simpler and more efficient compared to the conventional unsupervised learning approach~\cite{Hinton:2006dk, Tieleman:2008tj, Hinton:2012ig, Torlai:2016th}. 


To collect data and labels for the supervised learning task, we first run single bit-flip simulations of the model (\ref{eq:FKmodel}) to generate $50,000$ independent configurations together with the corresponding negative log-probabilities~\Eq{eq:FFK}. In light of the similarity between \Eq{eq:FFK} and \Eq{eq:FRBM}, we set $a_{i}=\beta U/2$ and focus on the fitting of $b_{j}$ and $W_{ij}$. Since sum of the softplus functions \Eq{eq:FRBM} is always positive, we subtract the minimum value in the collected labels to make them nonnegative. This again uses the fact that the fitting is only up to an additive constant. We use $80\%$ of the collected data for training and test on the remaining $20\%$ data to check the generalizability of the fitted neural net. In addition, we apply L2 regulation to the connection weights. This not only prevents overfitting but also makes the sampling of the RBM easier~\cite{Hinton:2012ig}. 

\begin{figure}[!t]
  \centering
  \includegraphics[width=\columnwidth]{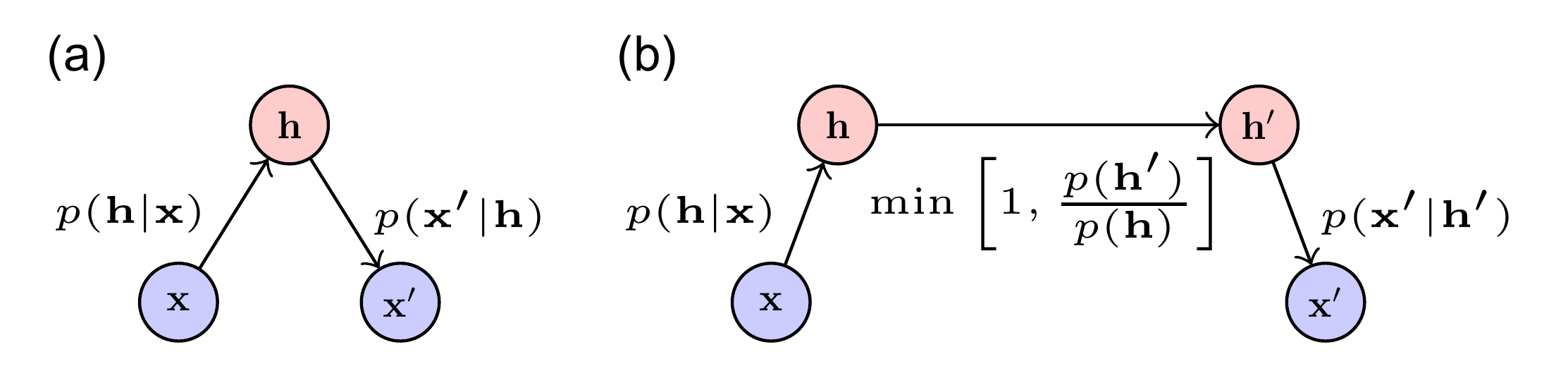}
  \caption{Two strategies of proposing Monte Carlo updates using the RBM. (a) The block Gibbs sampling. Given the visible variables $\mathbf{x}$, we sample the hidden variable $\mathbf{h}$ under the conditional probability $p(\mathbf{h}|\mathbf{x})$~\Eq{eq:phx}, then sample the visible variables under the conditional probability $p(\mathbf{x}'|\mathbf{h})$~\Eq{eq:pxh}. (b)
The block Gibbs sampling with an additional Metropolis step in between. It updates the hidden variable $\mathbf{h}\rightarrow\mathbf{h}'$ according to the log-probability of the hidden variables~\Eq{eq:Fh}. 
}
\label{fig:sampling}
\end{figure}

Figure~\ref{fig:fit} shows the negative log-probability of the test samples in solid red line and the predictions of the neural net in blue dots. The fitting successfully captures the overall trend of the physical probability distribution. Moreover, the connection weights $W_{ij}$ shown in Fig.~\ref{fig:weights} also acquire appealing physical meaning by acting as feature detectors for the visible variables. One clearly sees many cross structures corresponding to the staggered density-wave pattern of the localized fermions. There are also several features extend throughout the lattice, indicating that the hidden unit is sensitive to the nonlocal features of the physical model. These learned weights change with the temperature~\cite{SM}, showing that the RBM can pick up the characteristic features of the physics model automatically. When simulating the RBM as a statistical mechanics system, an activated hidden neuron will stimulate the corresponding features in the visible layer. This is similar to the application of RBM to images of hand-written digits. There, the RBM can pick up features like pen strokes and draw new images containing digits~~\cite{Hinton:2006dk, Tieleman:2008tj}. 

After training of the RBM, we use it to generate efficient Monte Carlo updates for the physical variables. To this end, we simulate the RBM using the Monte Carlo method and ensure the updates satisfy the detailed balance condition $\frac{T({\mathbf{x}}\rightarrow{\mathbf{x}'})}{T(\mathbf{x}'\rightarrow \mathbf{x})}= \frac{p(\mathbf{x}')}{p(\mathbf{x})}$ for the visible variables~\cite{SM}. Therefore, the Metropolis-Hastisings~\cite{Metropolis:1953in, Hastings:1970aa} solution of the acceptance ratio in \Eq{eq:dbl} reads 
\begin{equation}
A(\mathbf{x}\rightarrow \mathbf{x'}) = \min\left[1, \frac{p(\mathbf{x})}{p(\mathbf{x}')} \cdot \frac{p_\mathrm{FK}(\mathbf{x}')}{p_\mathrm{FK}(\mathbf{x})} \right]. 
\label{eq:metropolis}
\end{equation} 
Ideally, the acceptance ratio is one if the RBM fits the Falicov-Kimball model perfectly. In this case, one accepts all the proposals from the RBM as was attempted in the Ref.~\cite{Torlai:2016th}. However, in practice the fitting of the RBM is never perfect given the limited number of hidden units. Equation~(\ref{eq:metropolis}) corrects this error by rejecting unlikely proposals and guarantees \emph{exact} physical results even with an imperfectly trained RBM. 

A standard way to simulate the RBM is the block Gibbs sampling  approach. Because of the RBM's bipartite architecture, the conditional probability of the hidden variables factorizes $p(\mathbf{h}|\mathbf{x}) = p(\mathbf{x}, \mathbf{h})/p(\mathbf{x})=\prod_{j=1}^{M}p(h_{j}|\mathbf{x})$. Similarly, one has $p(\mathbf{x}|\mathbf{h}) 
=\prod_{i=1}^{N}p(x_{i}|\mathbf{h})$. And 
\begin{eqnarray}
p(h_{j}=1|\mathbf{x}) = \sigma\left(b_{j}+\sum_{i=1}^{N}x_{i}W_{ij}\right), \label{eq:phx} \\
p(x_{i}=1|\mathbf{h}) = \sigma\left(a_{i}+\sum_{j=1}^{M}W_{ij}h_{j}\right), \label{eq:pxh}
\end{eqnarray}
where $\sigma(z)= 1/(1+e^{-z})$ is the sigmoid function. 
The block Gibbs sampler samples back-and-forth between the hidden and visible layers using Eqs.~(\ref{eq:phx},\ref{eq:pxh}), shown in Fig.~\ref{fig:sampling}(a). When the simulation of the RBM is much cheaper than the original physical model, one can perform many of these block Gibbs sampling steps before evaluating~\Eq{eq:metropolis}. The RBM suggests nonlocal updates for the visible variables while still keeping the acceptance ratio high. 

Moreover, we argue that already a single Gibbs sampling step can be beneficial for the simulation of statistical physics model. Importantly, flipping a hidden variable with the  local Gibbs sampling may have nonlocal effects to the physical variables. This is because the hidden neuron may control an extended region of visible variables as shown in Fig.~\ref{fig:weights}. To further encourage this effect, one can perform additional sampling of the hidden variables in between the Gibbs sampling steps, shown in Fig.~\ref{fig:sampling}(b). We suggest to change the hidden variables $\mathbf{h}\rightarrow\mathbf{h}'$ and accept the update with  probability $\min[1, \frac{p(\mathbf{h}')}{p(\mathbf{h})}]$, where $p(\mathbf{h})=\sum_{\mathbf{x}}p(\mathbf{x},\mathbf{h})=e^{-F(\mathbf{h})}/Z$ is the ``free energy'' of the hidden variables 
\begin{equation}
-F(\mathbf{h})= \sum_{j=1}^{M} b_{j}h_{j}+\sum_{i=1}^{N}\ln\left(1+e^{a_{i}+\sum_{j=1}^{M} W_{ij}h_{j}}\right). \label{eq:Fh}
\end{equation}
Notice that the visible variables have been traced out in in \Eq{eq:Fh}, sampling according to $p(\mathbf{h})$ captures the generic distribution of the hidden features and is unaffected by the current visible variables. This further improves the sampling of RBM by avoiding the visible and hidden variables to lock each other's feature. 

\begin{figure}[t!]
  \centering
  \includegraphics[width=\columnwidth]{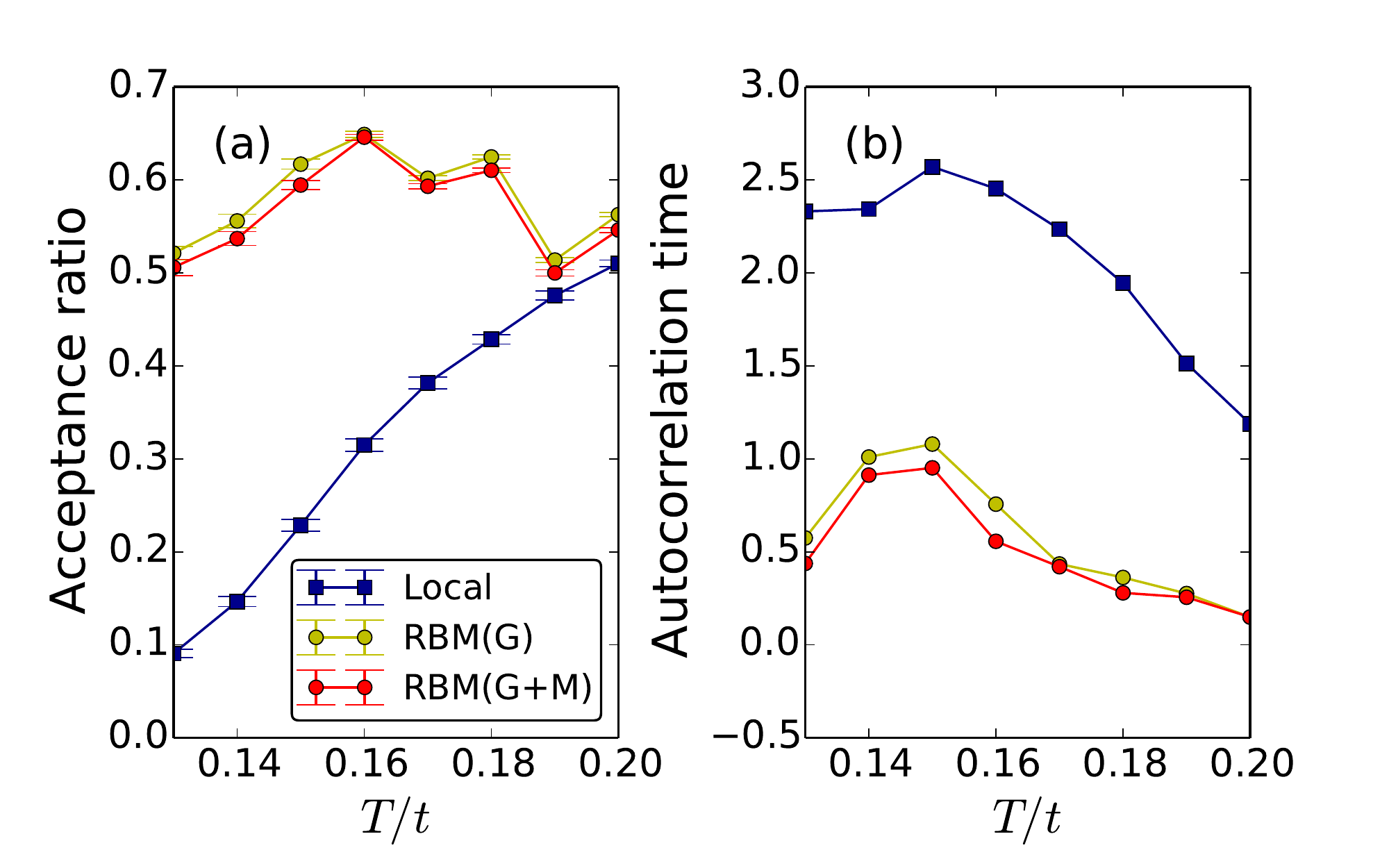}
  \caption{(a) The acceptance ratio and (b) the total energy autocorrelation time of the Falicov-Kimball model on a $N=8^{2}$ square lattice. Blue squares denote results of local bit-flip updates. The yellow and red dots are using the RBM update schemes of Fig.~\ref{fig:sampling}(a,b) respectively. The critical temperature is at $T/t\approx0.15$~\cite{Maska:2006id,Antipov:2014ij,PhysRevLett.117.146601}. The estimated physical observables agree within errorbars for all sampling approaches.}
  \label{fig:acc_tau}
\end{figure}

We demonstrate the improvement of using RBM in the Monte Carlo simulation in Fig.~\ref{fig:acc_tau}. The acceptance ratio of the single bit-flip updates simulation of the Falicov-Kimball model (\ref{eq:FKmodel}) decreases monotonously with the temperature lowers, shown in Fig.~\ref{fig:acc_tau}(a). This is because when the system enters into the CDW phase it is harder to add or remove the fermions. In contrast, the acceptance ratio of RBM updates remain high in the whole temperature range across the phase transition. This is because the RBM correctly captures the distribution of the physical system. As a better measure of the improvement, Fig.~\ref{fig:acc_tau}(b) shows the autocorrelation time measured in the unit of Monte Carlo steps per lattice site~\cite{Newman:1999fd}. The RBM updates reduce the autocorrelation time by at least a factor of two. The scheme of Fig.~\ref{fig:sampling}(b) with four additional bit-flip attempts of the hidden units further reduces the autocorrelation time. The overhead of performing the sampling using the RBM is $\mathcal{O}(MN)$, which is negligible compared to the cost of computing \Eq{eq:FFK} via diagonalizing the fermionic Hamiltonian. 

The proposed approach is general. Besides the Falikov-Kimball model and its relatives mentioned after \Eq{eq:FFK} it is straightforward to use the RBM in Monte Carlo simulations with binary degree of freedoms, such as the Ising and $Z_{2}$ gauge fields models, variational~\cite{VMC} and determinantal~\cite{Blankenbecler:1981vj} Monte Carlo simulation of the Hubbard models, and the Fermi bag approach of lattice field theories~\cite{PhysRevD.82.025007}. For models with continuous variables, one can use the RBM with Gaussian variables~\cite{Hinton:2012ig}. The RBM sampling approach can also be used in combination with the other efficient sampling approaches developed for statistical mechanics problems~\cite{duane1987hybrid, 1987PhRvL..58...86S, Wolff:1989iy, Prokofev:1998tc, 2001PhRvE..64e6101W, PhysRevLett.70.875, Motome:1999ig,  2001NuPhB.596..587A, Alvarez:2005cv,Kumar:2006iea}. 

To make the presentation cleaner, we divided the computational tasks into three phases: collecting the training data, fitting the RBM, and the actual Monte Carlo simulations. In future, one can use on-line learning to optimize the RBM progressively with newly collected configurations. After trained the RBM in the equilibration phase of the Monte Carlo simulation, one can use it to generate new samples with improved efficiency. One also needs to check the scalability of the proposed approach for larger and more complicated physical systems. 

Another future extension is to explore the deep Boltzmann machines~\cite{salakhutdinov2009deep} and deep belief nets~\cite{hinton2006fast} for Monte Carlo simulations. Deeper hierarchical structures may allow even higher level abstraction of the physical degree of freedoms. There were observations indicating that deep structure improves the mixing time of the Monte Carlo sampling~\cite{Bengio:2013ul}. To further exploit the translational symmetry of the physical problems, one may consider to use the shift invariant RBM~\cite{Anonymous:wjJtZlE6, Carleo:2016vp} or the convolutional RBM~\cite{Desjardins:2008wp, norouzi2009stacks, Lee:2009hj}. 

Last but not least, in this paper we view the RBM as a generative model and use it to accelerate the Monte Carlo simulation of physical systems. On the other hand, along the line of Ref.~\cite{Wang:2016tfa}, it is also interesting to view the RBM as an unsupervised feature detector and explore the patterns in the weights and the latent variables for discovery of physical knowledge.  

We end by noting an independent study~\cite{1610.03137}. It applies similar ideas to the Ising model. 

\paragraph{Acknowledgments--}
L.W. is supported by the Ministry of Science and Technology of China under the Grant No.2016YFA0302400 and the start-up grant of IOP-CAS. L.H. is supported by the Natural Science Foundation of China No.11504340. We acknowledge Jun-Wei Liu, Ye-Hua Liu, Zi-Yang Meng and Yang Qi for useful discussions. We use the Keras library~\footnote{Keras (\url{http://keras.io}) is a high level deep learning library based on Theano~(\url{http://deeplearning.net/software/theano}) and TensorFlow~(\url{https://www.tensorflow.org}).} for training of the neural network and the ALPS library~\cite{BBauer:2011tz} for the Monte Carlo data analysis. Our implementation of the restricted Boltzmann machine is based on the scikit-learn library~\footnote{\url{http://scikit-learn.org/stable/modules/generated/sklearn.neural_network.BernoulliRBM.html#}}. 

\bibliographystyle{apsrev4-1}
\bibliography{RBMpaper}

\clearpage 

\appendix
\section{Learned weights at $T/t=0.13$}
\begin{figure}[!t]
  \centering
  \includegraphics[width=\columnwidth]{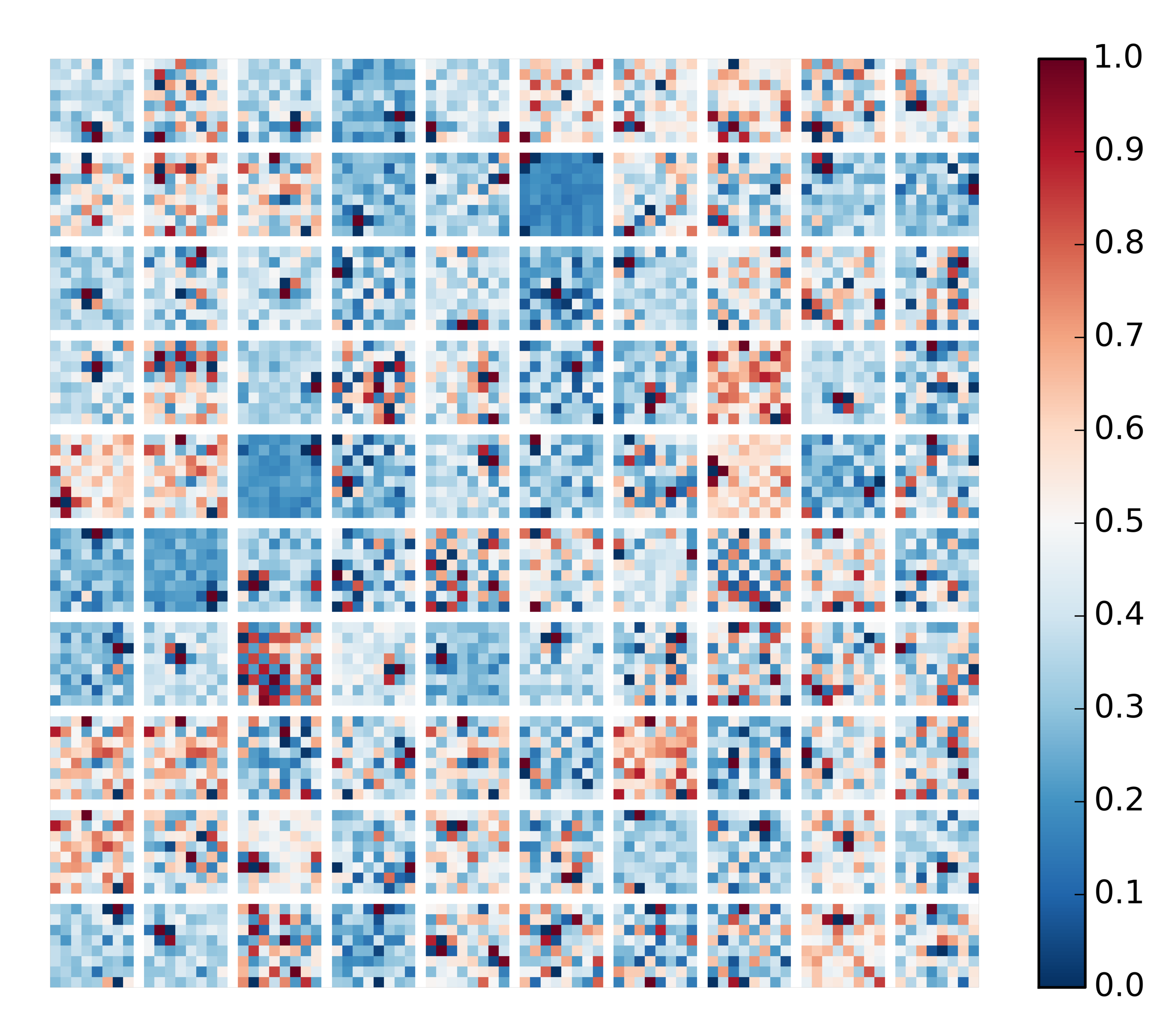}
  \caption{Connection weights $W_{ij}$ of the RBM at $T/t=0.13$. The other parameters are the same as the Fig.~\ref{fig:weights} of the main texts.}
  \label{fig:weights2}
\end{figure}

The learned weights change drastically near the critical temperature. Figure~\ref{fig:weights2} shows the weights learned by the RBM at lower temperature. Compared to Fig.~\ref{fig:weights} at $T/t=0.15$, there are more hidden neurons controlling extended regions of the visible variables, indicating enlarged correlation length at lower temperature. The checkerboard pattern of the low temperature phase is also more visible. 

\section{Proof of the detailed balance conditions}
We prove the simulation of the RBM shown in Fig.~\ref{fig:sampling} of the main texts satisfies the detailed balance condition.  

For the case of block Gibbs sampling shown in Fig.~\ref{fig:sampling}(a)
\begin{eqnarray}
  \frac{T (\mathbf{x} \rightarrow \mathbf{x}')}{T (\mathbf{x}' \rightarrow \mathbf{x})} &=& \frac{ p (\mathbf{h} | \mathbf{x} ) p (\mathbf{x}' | \mathbf{h}) }{
p (\mathbf{h} | \mathbf{x}' )p (\mathbf{x} | \mathbf{h} )} \nonumber \\ 
  & = & \frac{ p (\mathbf{x}, \mathbf{h}) p (\mathbf{x}',\mathbf{h})}{p (\mathbf{x}) p (\mathbf{h})} \cdot \frac{p (\mathbf{x}') p
  (\mathbf{h})}{p (\mathbf{x}', \mathbf{h}) p (\mathbf{x}, \mathbf{h})} \nonumber \\ & = & \frac{p (\mathbf{x}')}{p (\mathbf{x})}
\end{eqnarray}

For the case of Gibbs sampler with additional Metropolis steps for the hidden variables shown in Fig.~\ref{fig:sampling}(b)
\begin{eqnarray}
  \frac{T (\mathbf{x} \rightarrow \mathbf{x}')}{T(\mathbf{x}' \rightarrow \mathbf{x})} &=& \frac{p (\mathbf{h} | \mathbf{x} )  T (\mathbf{h} \rightarrow \mathbf{h}') p (\mathbf{x}' | \mathbf{h}'
  )}{p (\mathbf{h}' | \mathbf{x}' )
     T (\mathbf{h}' \rightarrow \mathbf{h}) 
     p (\mathbf{x} | \mathbf{h} ) } \nonumber \\ 
  & = & \frac{ p (\mathbf{x}, \mathbf{h}) p (\mathbf{x}',\mathbf{h}')}{p (\mathbf{x}) p (\mathbf{h}')} \cdot \frac{p(\mathbf{x}')p (\mathbf{h}) }{ p (\mathbf{x}', \mathbf{h}')p (\mathbf{x}, \mathbf{h})} \cdot \frac{p (\mathbf{h}')}{p (\mathbf{h})} \nonumber \\ & = & \frac{p (\mathbf{x}')}{p (\mathbf{x})}.
\end{eqnarray}
For the second equality we use that the Metropolis update of the hidden variable satisfies $T(\mathbf{h}\rightarrow \mathbf{h}')/T(\mathbf{h}'\rightarrow \mathbf{h})=p(\mathbf{h}')/p(\mathbf{h})$. This proof generalizes to compositions of several of such updates. 

\end{document}